\documentclass[9pt,twocolumn,twoside]{opticajnl}
\journal{opticajournal} 

\setboolean{shortarticle}{true}

\usepackage{lineno}
\usepackage{multirow}

\title{Laser written waveguides to the sample edge}

\author[1,*]{Zhi Kai Pong}
\author[1,*]{Mohan Wang}
\author{Ji Qin}
\author[1]{Martin J. Booth}
\author[1,2]{Patrick S. Salter}

\affil[1]{Department of Engineering Science, University of Oxford, Parks Road, Oxford OX1 3PJ, UK}
\affil[*]{These authors contributed equally to this work}
\affil[2]{patrick.salter@eng.ox.ac.uk}

\begin{abstract}
A method is presented for fabrication of femtosecond laser written waveguides in glass to remove the need for polishing of substrates after processing. It is shown that by amplitude masking the fabrication laser beam near the sample edge and increasing the pulse energy it is possible to write waveguides that are not affected by edge aberrations and display mode profiles well matched to single mode fiber. Results are presented for different depths in fused silica and borosilicate glass substrates. The transmission from fiber to photonic circuit is significantly improved for situations where it is not possible to polish glass substrates after laser writing, creating new opportunities in photonic packaging.
\end{abstract}

\setboolean{displaycopyright}{false} 

\begin{document}

\maketitle


Laser writing inside transparent materials with ultrashort pulses~\cite{Gattass2008} offers incredible versatility for three dimensional internal structuring. Laser inscription of waveguides in glass represents an area of much research activity and has progressed significantly since early demonstrations~\cite{Davis96, 10.1063/1.120327}. The capability for 3D networks of waveguides creates opportunities missing from other fabrication methods~\cite{Osellame_review, ZHONG2025100157}. Highly sophisticated devices are now being reported for a myriad of applications including quantum photonics~\cite{Sansoni, Meany}, topological photonics~\cite{ Szameit}, astrophotonics~\cite{jovanovic, Arriola:14} and biosensing~\cite{B808360F}. Likewise photonic interconnects for data communication are drawing much attention from industry~\cite{Djogo_2019, 10494504, Fernandez}.

Photonic circuits in glass typically need to be interfaced with external optical fiber connections, typically via butt coupling at the sample edge. The laser write process involves translation of the glass substrate relative to a fixed laser focus, tracing out the optical waveguide in three dimensions. However, often it is difficult to laser write waveguides near the edge of the substrate due to optical aberrations, since the focusing cone of the fabrication laser intersects both the top and side face of the substrate~\cite{salter2012}. Therefore, it is common after laser writing of a photonic circuit to polish back the sides of the glass substrate to reveal the section of waveguide unaffected by the edge aberration, for efficient butt coupling to the external optical fiber source or sink. In this letter, we present an optical fabrication technique that circumvents edge aberrations, allowing the laser writing of waveguides directly to the sample edge without the need for polishing post-processing.  

\begin{figure}[ht]
\centering
\fbox{\includegraphics[width=\linewidth]{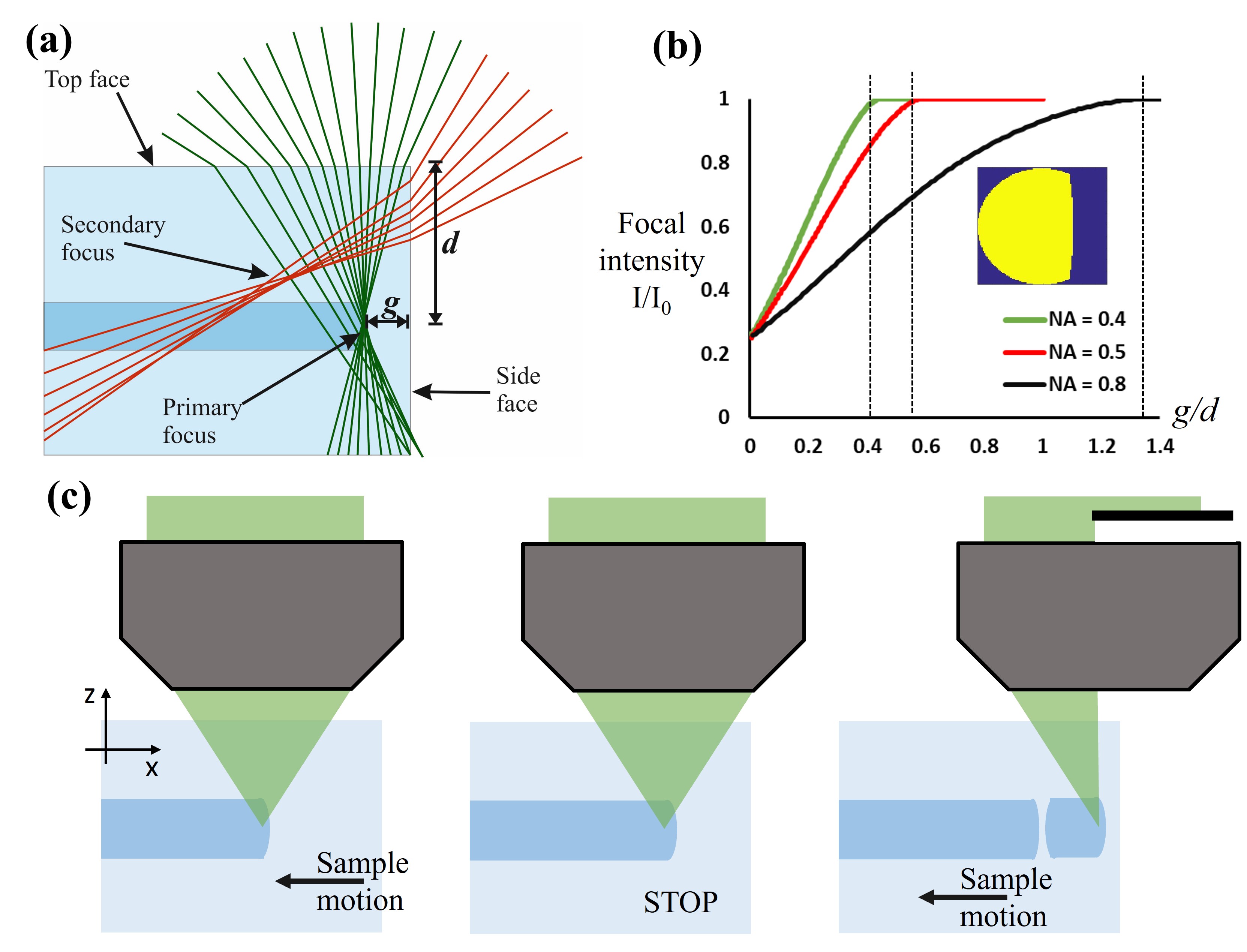}}
\caption{(a) Ray tracing showing refraction of the fabrication laser near the edge of the glass substrate, with green rays intersecting the top face of the substrate while red rays are incident on the side face. (b) Reduction of the peak focal intensity as a function of depth $d$ and distance  from the sample edge $g$. Results are shown for different NA lenses. Inset shows an example plot of the objective lens pupil plane where only regions in yellow contribute rays that are incident on the top surface of the substrate and included in the focal intensity calculation. (c) Schematic for writing waveguides to the edge. The bulk waveguide is written as normal (left), and intentionally terminated some distance from the substrate edge (centre). The fabrication laser pupil plane is masked to prevent rays intersecting the edge face and the remaining segment of the waveguide to the sample edge is written (right).}
\label{fig:false-color}
\end{figure}

The difficulty of writing waveguides near a sample edge is illustrated in Fig. 1(a). Near the sample edge, rays of light from the fabrication laser are refracted by both the top and side surfaces of the substrate, which leads to a spatial and temporal splitting of the focus formed from the rays incident on each face~\cite{salter2012}. This edge aberration reduces the pulse energy at the primary focus (resulting from rays incident on the top surface), causing the waveguide to terminate before it nears the edge of the sample. The variation in focal intensity as a function of the gap $g$ from the sample edge is explored in Fig 1(b) for different depths $d$ and focusing numerical aperture (NA). It is considered that only the rays incident on the top surface contribute to the focal intensity $I$, and that the \textcolor{black}{boundary} in the pupil plane separating the top and side rays is given by~\cite{salter2012}:

\begin{equation}
\rho = \frac{1}{\text{NA} \sqrt{1+(\frac{d}{g})^2\cos^2{\theta}}}
\label{eq:refname1}
\end{equation}

Where $\rho$ and $\theta$ are the normalized radial and azimuthal coordinates in the pupil plane. \textcolor{black}{More details on the simulation in Fig 1(b) are provided in the Supplementary Information (SI) Fig. S1} It is seen that as the gap $g$ decreases the normalized focal intensity ($I/I_0$) drops to a minimum of a quarter when located directly at the substrate edge. At this point, not only is half the incident light not effectively contributing to the focus, as it is refracted by the side interface, but also the associated amplitude aberration doubles the focal area due to the reduction in effective NA, leading to the factor of four reduction in focal intensity. Also noticeable is that when using higher NA lenses, the edge aberration becomes prominent further from the edge of the substrate, due to the higher angle focusing and larger beam area at the top surface. Similarly, at greater depths inside the substrate the edge aberration reduces the focal intensity at greater distances from the edge. As an example, for a typical focusing NA of 0.5 and at a depth of 200 $\mu$m in glass, the focal intensity is reduced to 0.75$I_0$ at a distance of 70$\mu$m from the edge and 0.5$I_0$ at 40$\mu$m. The waveguide fabrication is consequently drastically affected in this region, which has a significant effect on transmission when butt-coupling with external optical fiber; hence, an additional polishing step is required.

In some situations, it would be beneficial to remove any form of post processing such as polishing, if glass chips are already bonded to other components. Therefore, new manufacturing methods are needed to facilitate waveguide fabrication closer to the substrate edge. \textcolor{black}{Previously, we showed that the edge aberration can be fully compensated using adaptive optics~\cite{salter2012}, although the limited refresh rate of a liquid crystal spatial light modulator (SLM) makes it prohibitively slow and incompatible with standard waveguide fabrication that occurs at speeds of millimeters per second. Here we propose a simpler scheme, illustrated in Fig. 1(c) which is better suited to device manufacturing.} When the laser focus is a set distance from the interface of interest, the waveguide is intentionally terminated. An amplitude mask is then applied to the fabrication laser to block one half of the pupil plane and prevent any rays from intersecting with the side facet of the sample as the gap to the edge of the sample is further reduced. Concurrently, the laser power is increased to ensure there is enough pulse energy at focus for the desired fabrication. The amplitude mask is an important feature in order to avoid fabrication in the secondary focus as the pulse energy is raised and to stop ablation of the side face of the substrate.  The remaining segment of the waveguide in the edge region is then written to connect the bulk waveguide up to the edge of the sample.

To implement the new method for writing up to the edge of the substrate, we used the fabrication system outlined in detail in the SI Fig. S2 and Ref~\cite{sunlpr2024}. In short, the beam from an ultrafast laser (Pharos SP6, 515nm, 1MHz, 170fs) was expanded on to a SLM (Hamamatsu X10468). The SLM was relayed in a $4f$ configuration to the back aperture of an objective lens (Zeiss 20$\times$ 0.5NA). The samples were mounted on air bearing translation stages (Aerotech ABL10100) for transverse waveguide writing. For amplitude masking of the beam,  we used the method described in Ref~\cite{Salter:12} by applying a blazed grating to the SLM. A pinhole of diameter 1~mm was placed in the Fourier plane of the SLM and positioned so as to block the zero order beam. The amplitude of the beam at the back aperture of the objective lens could thus be controlled by selective application of the blazed grating on the SLM. We note that the SLM is not the only way to achieve amplitude masking and an actuated opaque plate placed before the objective lens can still provide a successful implementation of the technique for writing to the edge. In our system, however, the SLM was also used for aberration correction during waveguide fabrication, which is particularly relevant when writing waveguides at different depths within the substrate~\cite{Mauclair:08, Bisch2018}.

\begin{figure}[ht]
\centering
\fbox{\includegraphics[width=7cm]{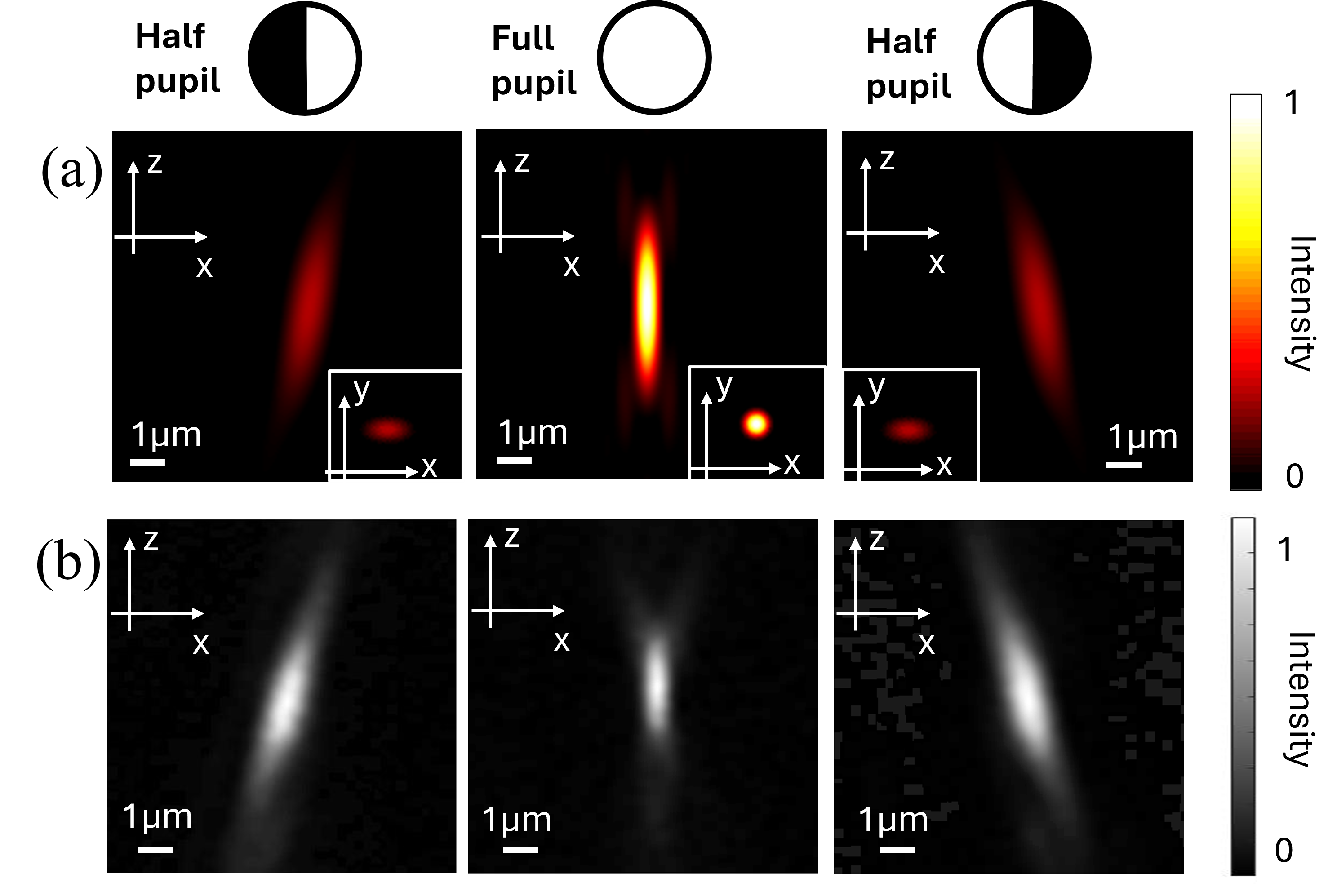}}
\caption{Distribution of the laser fluence at focus from theory and experiment for different levels of amplitude masking in the pupil plane of the objective lens. (a) shows theory for the axial intensity distributions and inset the transverse profile. (b) shows experimentally measured plots of the focal fluence, with adjusted laser power to aid visualisation.}
\label{fig:false-color}
\end{figure}

To evaluate the effect of the amplitude mask on the laser fabrication focus, images are presented in Fig. 2 from the experimental system and simulation. To acquire experimental images of the focal fluence, a mirror specimen was placed in the focal plane of the objective lens and the reflected signal imaged onto a CCD. Axial translation of the mirror provides z-sectioning of the laser focus (Fig. S3). Simulated images were calculated based on a Fourier optics representation of the objective lens, with phase and amplitude control over the field in the back aperture. There is good agreement between the simulated and experimental images in Fig. 2. With half the pupil amplitude blocked there is slight distortion of the laser focus relative to the diffraction limit, notably generating a tilted intensity distribution in the direction normal to the substrate edge. The reduction in effective NA causes an increase in focal width by a factor of two along this same direction. However, the lateral extent of the focus in the plane parallel to the substrate is unaffected, likewise the full range of zenithal angles transmitted by the lens means there is little change to the axial extent of the focus. Fig. S4 additionally explores the effects of amplitude masking on the focal fluence with different amounts of depth dependent phase aberration.

For the experiments, two different glass substrates were used, both of which are important materials for laser written waveguides: fused silica (FS) and Corning Eagle 2000 (E2K). \textcolor{black}{Care needs to be taken during polishing for substrate preparation to avoid chips at the edge of the glass, or chamfers of width greater than several micometer, which are detrimental to the focusing.} To form the waveguides in the bulk of the sample, we used  single scan transverse writing at a laser pulse repetition rate of 1MHz with a feed rate of 4~mm/s and a pulse energy of 50~nJ  for FS, and 6~mm/s and 80~nJ for E2K respectively. Waveguide writing parameters were used from previous studies, which looked to minimize insertion loss for waveguides written in the bulk of these materials~\cite{sunlpr2024}. After optimization to match the mode profile of the bulk waveguide while keeping the feedrate fixed at 4~mm/s for FS and 6~mm/s for E2K), pulse energies for writing the edge waveguide segment were determined as  120~nJ for FS and 214~nJ for E2K (measured before the pupil mask). We note that the increase in pulse energy by a factor of $\approx$2.5 for both materials is significantly less than the factor 4 predicted by theory to match the peak focal intensity. This indicates that the waveguide formation process is not completely driven by peak intensity and, as is often the case, the optimum pulse energy needs to be found through experimental tests, where the main criterion is matching the guided mode profile of the waveguide segment with that of the bulk waveguide.

\begin{figure}[ht]
\centering
\fbox{\includegraphics[width=7.5cm]{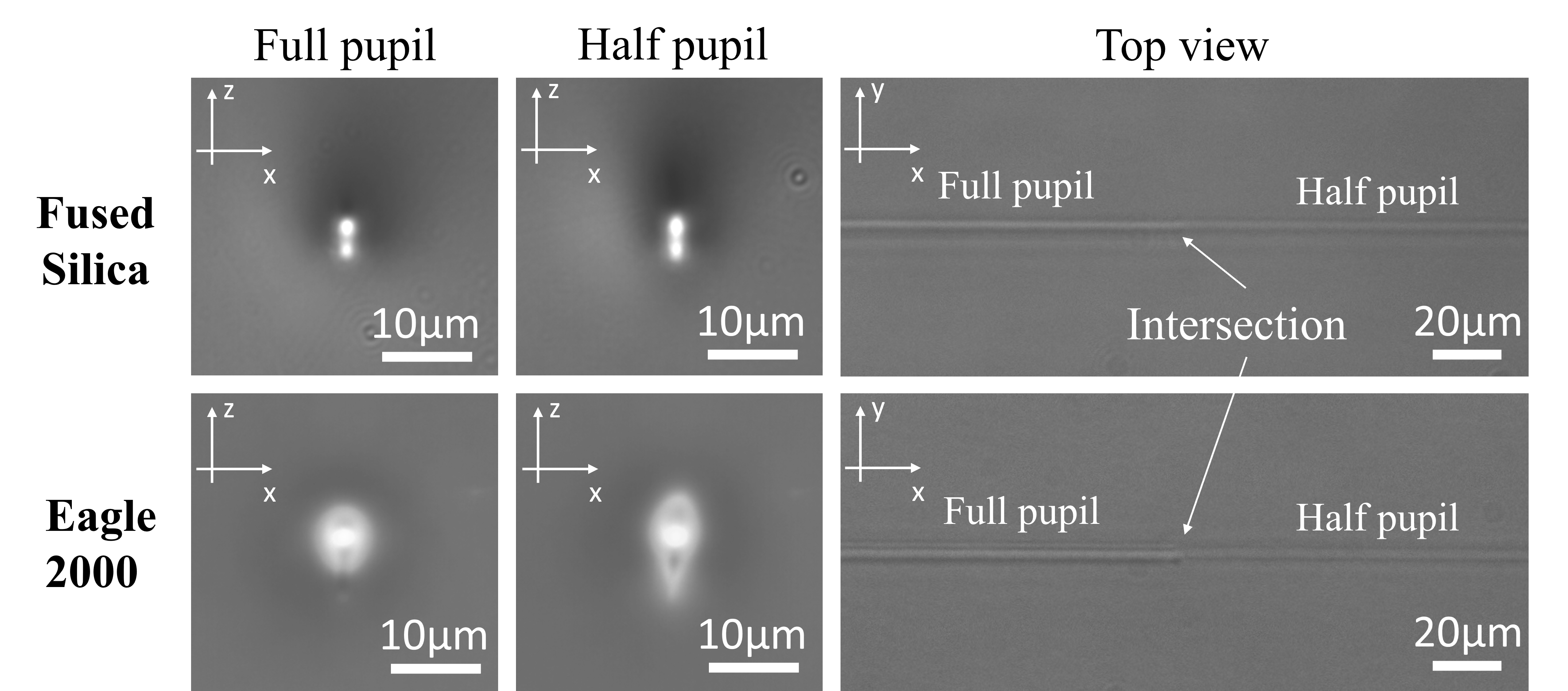}}
\caption{Cross-section and top view images of the full pupil and half pupil waveguides in FS and E2K.}
\label{fig:false-color}
\end{figure}

Fig. 3 shows transmission optical microscope images of the cross-section and top-view for both the full pupil and half pupil waveguides. In FS, the waveguides look almost identical in cross-section, \textcolor{black}{with physical dimensions measured from the transmission microscope image as 3.4~$\mu$m $\times$ 5.9~$\mu$m for the full pupil and 3.3~$\mu$m $\times$ 5.8~$\mu$m for the half pupil. In addition, the waveguides appear identical in the top-view image either side of the intersection between edge and bulk waveguide segments. In E2K, there is a noticeable difference in appearance between the edge and bulk waveguide, which is particularly clear in the cross-section image (physical dimensions of 7.1~$\mu$m $\times$ 8.1~$\mu$m versus 5.9~$\mu$m $\times$ 9.7~$\mu$m).} However, we note that quite different refractive index structures in laser written waveguides can give rise to very similar profiles of the associated guided mode~\cite{sunlpr2024}, and it is the mode which should be consistent between the waveguide segments to minimize the coupling loss.

Fig. 3 additionally shows the intersection point between the edge and bulk waveguide segments. When starting and terminating waveguide segments inside the substrate, the laser was gated on/off at constant feedrate to avoid an increased laser dose at the intersection points. A gap of 2~$\mu$m was used between adjacent waveguide segments to minimize separation whilst avoiding overlap. The microscope images in Fig.3 show no indication of additional defects caused at the intersection point.

\begin{figure}[ht]
\centering
\fbox{\includegraphics[width=7.5cm]{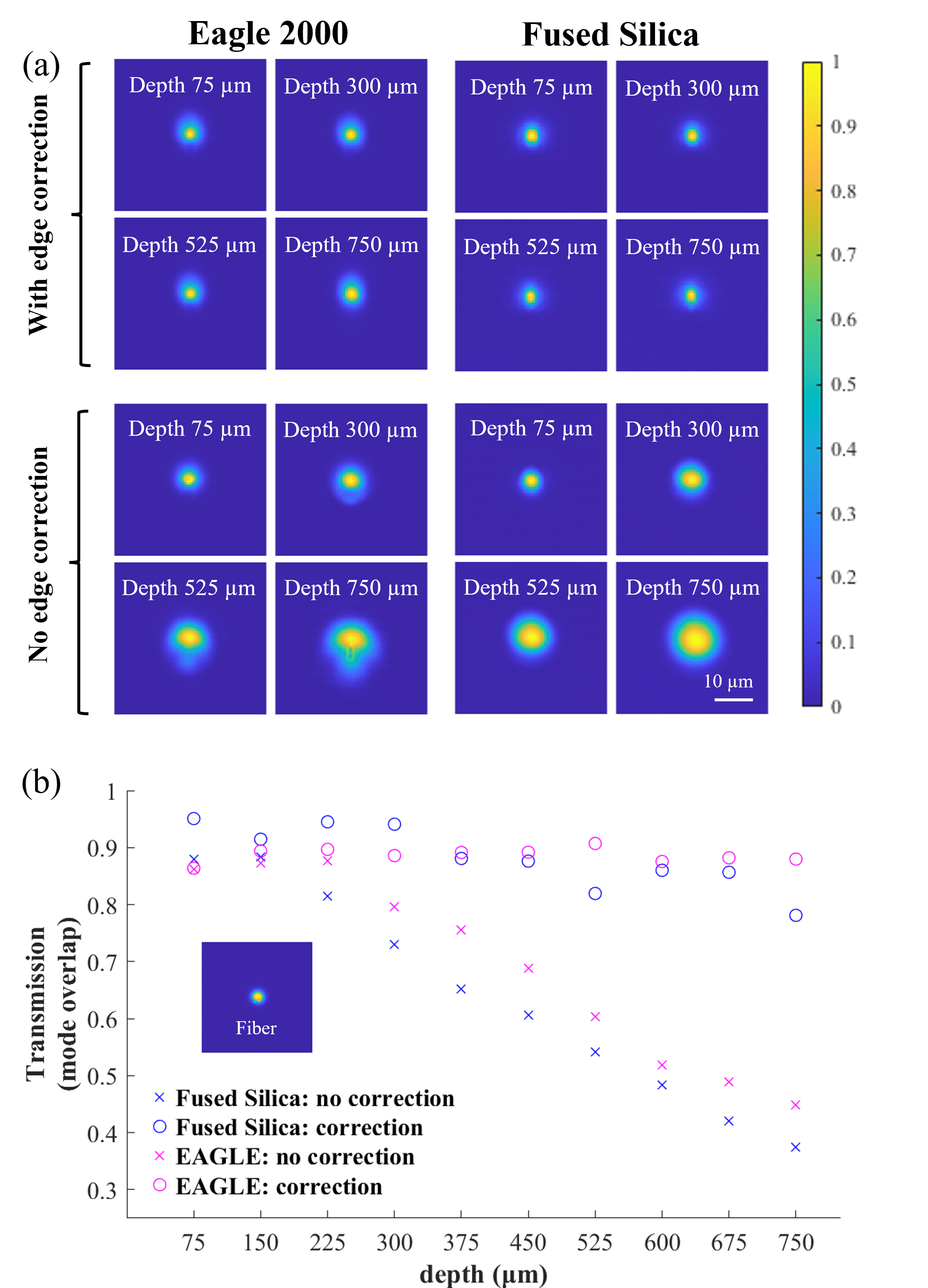}}
\caption{(a) Example images of the experimentally measured beam profile ($\lambda$=780~nm)  at the output face of the glass chip for waveguides at different depths in E2K and FS, laser written with and without the edge correction technique. (b) Calculated transmission into butt-coupled optical fiber based upon an overlap integral between the measured optical beam profile and that of the the measured fiber mode (inset). Scale bars are consistent in all images. }
\label{fig:false-color}
\end{figure}

\begin{table}
    \centering
    \begin{tabular}{c|c|c|c|c|c|c}				
        & Depth ($\mu$m) & 150 & 300 & 450 & 600 & 750  \\
        \hline
         \multirow{2}{*}{I} & E2K (dB) & 0.2 & 0.7 & 1.1 & 2.1 & 3.6\\	 	
         & FS (dB) &  0.4 & 1.2	 & 1.9	 &  2.7	 & 3.3\\
        \hline
         \multirow{2}{*}{EL} & E2K (dB) & 1.3 & 1.2 & 1.1 & 1.3 & 1.4\\
         & FS (dB) &  0.3 & 0.0	 & 0.4	 &  0.5	 & 0.1\\
         
    \end{tabular}
    \caption{The insertion loss improvement (I) for waveguides butt-coupled to optical fiber at one interface as a function of depth by using the edge correction technique, and the extra loss (EL) of such waveguides compared to a regular full pupil waveguide with polished end facets.}
    \label{tab:my_label}
\end{table}

 Waveguides were fabricated at different depths within the substrate, to explore different amounts of edge aberration (as predicted in Fig. 1b). All depth-related spherical aberrations induced by refraction at the top interface were corrected using the SLM~\cite{Bisch2018}. \textcolor{black}{In each case, the waveguide fabrication was stopped at $g=10~\mu$m from the edge of the substrate by gating the laser emission while maintaining a constant feed rate; this was found to bring the waveguide sufficiently close to the substrate edge while avoiding any potential ablation of the edge facet.} Directly following fabrication, without any polishing of the substrate, light at a wavelength of 780~nm was coupled into the waveguides and Figure 4(a) presents  measurements of the resultant beam profile imaged at the output face of the substrate, which corresponds to the intensity profile that would be encountered by a butt-coupled optical fiber.  

As can be seen in Figure 4(a),  for both E2K and FS substrates the measured beam size increases substantially with depth in the substrate (with additional analysis in Fig. S5). This is to be expected, since at greater depths the edge aberration leads to reduction of the fabrication focal intensity, and hence waveguide termination, further from the substrate edge. Therefore, the beam imaged at the output facet has already started to diverge from the end of the waveguide, \textcolor{black}{whilst noting that the divergence is significantly weaker than in air, due to the reduced numerical aperture of the waveguide within the glass~\cite{Qin:26}.} For E2K, there is additionally a strong distortion of the beam profile at greater depths for the case without edge correction, since the heat accumulation regime causes a gradual taper at the waveguide termination, which is further influenced by the focal split from the edge aberration. When the edge correction technique is applied (using a half pupil illumination and raised laser pulse energy), a constant beam size at the end facet is maintained at all depths, since the influence of the edge aberration is avoided and waveguides can be written directly to the substrate edge. Furthermore, the beam profile measured for the case of the edge correction is well matched to the mode of the input optical fiber (which is shown as an inset in Fig 4b) at all depths. 

 By performing a mode overlap integral for the beam profile measured at the output face with that of the mode of the coupling optical fiber, an estimate of the coupling loss can be calculated, as shown in Fig. 4(b). With the edge correction technique, the predicted coupling loss is relatively uniform and below 1dB. Without the edge correction technique, coupling loss below 1dB can still be achieved at shallow depths below 200~$\mu$m where the edge aberration is less significant, but at greater depths the coupling loss steadily increases to exceed 3dB at depths of 600~$\mu$m.

The waveguides at different depths were butt-coupled to single mode optical fiber and probed at a wavelength of 780~nm to experimentally determine the improvement in transmission from the edge correction technique. The optical fiber was coupled at one interface while the light was collected free space from the other interface and imaged onto a power meter. Table 1 presents the improvement in transmission from the edge correction technique when compared to standard fabrication through the length of the substrate at fixed pulse energy and full pupil illumination. It is seen that at all depths the edge correction technique brings benefit, but particularly so for depths of 300~$\mu$m and above, in agreement with the analysis of Fig. 4(b).  

The total insertion loss of the waveguides written using the half-pupil edge writing method over a 20~mm length glass sample was measured to be 2.6~$\pm 0.23$~dB for FS and 2.2~$\pm 0.1$~dB for E2K. This includes waveguide propagation loss, Fresnel loss, and coupling losses between the fiber and the half-pupil waveguide, as well as between the half-pupil and full-pupil waveguides. The measured values are independent of waveguide depth over the range of 75 to 750 µm due to the correction of depth induced aberrations. The samples were then polished back to remove the edge-written region  and compare with the total insertion loss of a standard polished sample, which was found to be 2.3~$\pm 0.21$~dB for FS and 1.0~$\pm 0.06$~dB for E2K. The difference between the as-written and polished back samples is characterized as the extra loss (EL) from the edge write technique and is also given for each depth in Table 1. It is seen that for FS the edge write technique introduces no additional loss, indicating very good mode matching between waveguide segments. In E2K, the EL is consistently around 1db across all depths, due to difficulty in accurately matching waveguide mode profiles in the cumulative heating regime. The propagation loss was also measured using the cut-back method as 0.8dB/cm and 0.7dB/cm for the full pupil and half pupil in FS, and <0.1dB/cm for both the full pupil and half pupil in E2K, again independent of the waveguide depth. This demonstrates that amplitude masking in the pupil of the objective to avoid edge aberrations does not affect the ability to laser write low-loss waveguides.

In summary, we have demonstrated a new technique for the fabrication of laser written waveguides in glass up to the edge of the substrate without post-processing. Through amplitude masking of the fabrication laser beam and increasing the write pulse energy, it is possible to circumvent negative effects associated with the optical aberration which is present near the sample edge. The results showed enhanced transmission to butt-coupled optical fiber. The new technique can be advantageous in scenarios where it is not practical to polish glass substrates after laser fabrication and could enable more effective scale-up of photonic chip manufacturing. It is also quite general in application, relevant for other materials, including crystals, and other devices that are fabricated by internal laser processing.


\begin{backmatter}
\bmsection{Funding} Engineering and Physical Sciences Research Council (EP/X017931/1, EP/W025256/1). Advanced Research and Innovation Agency (ARIA) NACB-SE02-P01.





\bmsection{Disclosures}  MB: Opsydia Ltd (I,C,S). PS: Opsydia Ltd (I,C,S).

\bmsection{Data availability} Data underlying the results presented in this paper are not publicly available at this time but may be obtained from the authors upon reasonable request.

\bmsection{Supplemental document}
 See Supplement 1 for supporting content.
\end{backmatter}


\bibliography{sample}

\bibliographyfullrefs{sample}

\end{document}